\documentclass{IEEEtran}
\usepackage[T1]{fontenc}
\usepackage[latin9]{inputenc}
\usepackage{xcolor}
\usepackage{amsmath}
\usepackage{graphicx}
\usepackage{esint}
\PassOptionsToPackage{normalem}{ulem}
\usepackage{ulem}

\makeatletter

\providecommand{\tabularnewline}{\\}
\providecolor{lyxadded}{rgb}{0,0,1}
\providecolor{lyxdeleted}{rgb}{1,0,0}

\DeclareRobustCommand{\lyxdeleted}[3]{{\color{lyxdeleted}\lyxsout{#3}}}
\DeclareRobustCommand{\lyxsout}[1]{\ifx\\#1\else\sout{#1}\fi}

\makeatother

\begin{document}
\title{Derivation of the Antenna Contribution to the Reverberation-Chamber
$Q$-Factor Based on Antenna Scattering-Matrix Theory}
\author{\author{Julien de Rosny, \IEEEmembership{Senior Member, IEEE}, Youssef Rammal, Isma\"{i}l Ahmed Bouha and Fran\c cois Sarrazin 
\thanks{Manuscript created May 4, 2026.} 
\thanks{Julien de Rosny and Youssef Rammal are with ESPCI Paris, PSL Research University, CNRS, Institut Langevin, Paris, France.}
\thanks{Isma\"{i}l Ahmed Bouha and Fran\c cois Sarrazin are with Univ Rennes, INSA Rennes, CNRS, IETR-UMR 6164, F-35000 Rennes.}
\thanks{This work was support in part by the French ``Agence Nationale de la Recherche" (ANR) under Grant ANR-22-CPJ1-0070-01.}}}
\maketitle
\begin{abstract}
A radio antenna is primarily designed to convert electromagnetic waves
into electrical current and vice versa. However, a part of the incident
wavefield is scattered due to structural effects andreflection at
the antenna's electrical port. Because the reflected power depends
on the load impedance, an antenna can also be referred to as a loaded
scatterer. Its interaction with electromagnetic waves is characterized
by absorption and scattering cross-sections (ACS and SCS). When immersed
in a diffuse field, such as the one generated within a reverberation
chamber (RC), the impact of the loaded antenna is determined by averaging
these properties over incident angles. Of particular interest is the
averaged ACS from which one can derive the antenna contribution to
the RC quality factor (Q-factor). Current formulations rely on different
power budget analyses which do not account for wave interferences
between the ingoing and outgoing fields. Moreover, existing formulations
consistently neglect the structural component. In this paper, we introduce
a rigorous formulation of the antenna contribution to the RC Q-factor
which takes into account the aforementioned effects. The antenna is
modeled using the scattering-matrix theory, which linearly links the
ingoing and outgoing waves in terms of spherical harmonics expansion.
The derived theory is validated using several numerical simulations
based on a Method-of-Moment code. The model's ability to retrieve
antenna properties from multiple Q-factor estimations in an RC is
then demonstrated. All results are compared with \lyxdeleted{Francois SARRAZIN}{Mon Apr  7 11:49:36 2025}{ }existing
formulations.
\end{abstract}

\section{Introduction}

\IEEEPARstart{R}{everberation} chambers (RCs) are now widely used in a broad range of applications, including electromagnetic compatibility testing \cite{Besnier2013book} and antenna characterization. The latter encompasses not only average quantities, such as radiation efficiency \cite{Holloway2012,Krouka2022biased}, but also line-of-sight contributions, including radar cross-section (RCS) \cite{Reis2021,Soltane2018} and gain patterns \cite{Lemoine2013,Garcia2014,Soltane2020}. An important parameter for characterizing an RC is its quality factor ($Q$-factor), which describes the chamber's ability to store energy. This $Q$-factor is often referred to as a composite $Q$-factor, as it can be decomposed into several loss mechanisms, including wall losses, losses due to loaded objects, aperture losses, and antenna losses, each corresponding to a specific contribution to the overall RC $Q$-factor \cite{Hill1998}. In particular, antennas are essential for any electromagnetic measurement, and their contribution to the overall losses can be significant, especially at lower frequencies \cite{Loughry1991,Genender2010}. Therefore, it is important to develop models that accurately describe their absorbing properties in such environments. Moreover, such models are valuable for enabling noninvasive antenna characterization within RCs \cite{Krouka2022,Galesloot2023}, avoiding the use of cables that can disturb the impedance and radiation properties of the antenna under test \cite{Icheln1999,Huitema2014}.

An antenna illuminated by incoming electromagnetic waves can be viewed as a loaded scatterer, i.e., an antenna terminated by a passive load at its electrical port. From a general perspective, the interaction of a scatterer with an incident wave can be characterized through the scattering and absorption cross-sections (SCS and ACS, respectively). In a large RC, the interaction of the scatterer with the diffuse field--i.e., a random superposition of incident waves arriving from many different directions--is described by the averaged SCS (ASCS) \cite{Lerosey2007} and the averaged ACS (AACS) \cite{Carlberg2004}, obtained by averaging over all possible angles of incidence. Rather than relying on the AACS, the absorption of an antenna in an RC is predominantly characterized by its contribution to the RC $Q$-factor, which is inversely proportional to the AACS and directly measurable. A first model of antenna-related losses within RCs was derived by D.~Hill in 1998 \cite{Hill1998}. However, as pointed out in \cite{Cozza2018}, this model assumes that losses arise solely from the power dissipated in the antenna termination load. It was later refined in 2018 \cite{Cozza2018} to better account for dissipation associated with the antenna radiation efficiency. Nevertheless, both models neglect the \emph{structural} component of the antenna absorption properties. Indeed, following pioneering studies on antenna scattering in the mid-20th century \cite{Dicke1947,King1949}, it has been well established that the field scattered by an antenna consists of two contributions: the \emph{structural mode} and the \emph{antenna mode} (also referred to as the radiation mode) \cite{Hansen1989,Harrington1964}. On the one hand, the structural mode describes the behavior of the object independently of its function as an antenna. Fundamental considerations imply that this structural component depends on a reference, as it is defined as the response of the antenna under an arbitrary load condition. While short-circuit, open-circuit, and matched-load references have all been proposed (and are equally valid), the most widely used formulation relies on the conjugate matched load introduced by R.~Green \cite{Green1963}. On the other hand, the antenna mode is directly related to the antenna's radiation and impedance properties, as well as to the chosen reference load. This decomposition has been extensively studied in the antenna scattering community and has found applications not only in antenna RCS reduction \cite{Knott2004}, but also in antenna measurements, where antenna parameters can be retrieved from backscattering measurements, enabling contactless characterization \cite{Reis2021,AppelHansen1979,Wiesbeck1998}.

The objective of this paper is to introduce a complete and rigorous model of the antenna contribution to the RC $Q$-factor, accounting for both structural and antenna modes. To achieve this, the antenna is represented using a scattering matrix \cite{Dicke1947,Collin1969} where the incoming and outgoing waves are defined according to the spherical wave expansion (SWE) \cite{Hansen1988}. The antenna's contribution to the RC $Q$-factor is then derived by computing the AACS of an antenna immersed in a diffuse field. This new derivation highlights the existence of not only structural and antenna modes but also their interference, which cannot be modeled through a power budget analysis only. This model will provide a better understanding of the loss mechanisms induced by an antenna within an RC and may also be valuable for antenna characterization based on RC $Q$-factor estimations.

The paper is structured as follows. Section~\ref{sec:HillCozza} briefly reviews the current formulation of the antenna's contribution to the RC $Q$-factor based on a power budget analysis and their inherent limitations. Then, Section~\ref{sec:newmodel} introduces the new formulation starting from the $Q$-factor definition in terms of AACS and deriving it from the antenna scattering-matrix representation. Section~\ref{sec:comparison} provides a discussion of the inherent assumptions made in the previous models. Section~\ref{sec:Validation} presents a numerical validation of the introduced model, based on the Method of Moments by considering a half-wavelength dipole antenna of various radiation efficiencies. Then, Section~\ref{sec:Antenna-characteristics-retrieva} exhibits the capability to retrieve antenna radiation efficiency and input impedance from RC $Q$-factor estimations for multiple load conditions. Finally, Section~\ref{sec:fullwave} presents a second numerical validation based on full-wave simulations of a realistic patch antenna.

\section{Current formulations of the antenna contribution to the RC $Q$-factor}

\label{sec:HillCozza}

\begin{figure}
\centering{}\includegraphics[width=0.8\columnwidth]{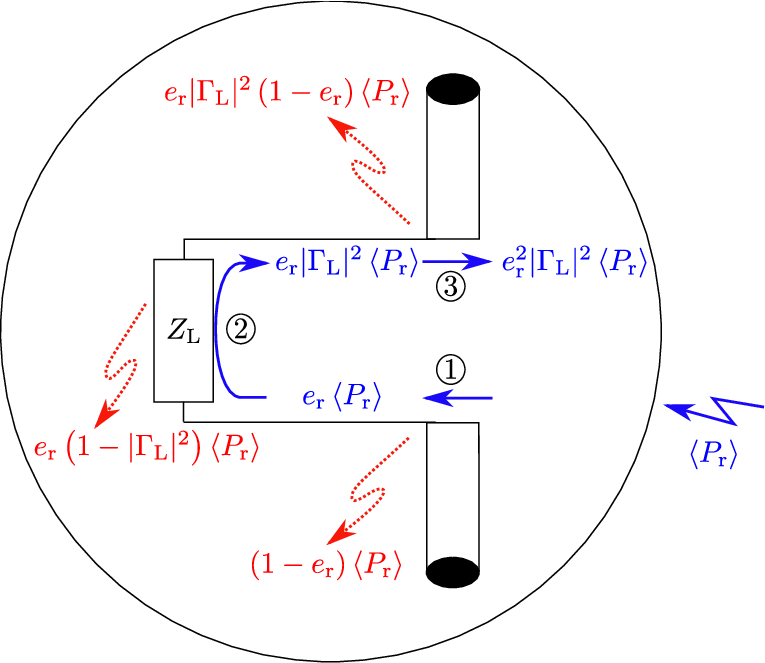}\caption{Description of the power dissipation process of a loaded scatterer
as stated in \cite{Hill1998} and \cite{Cozza2018}. Red color indicates
power losses whereas blue color indicates transmitting power. \label{fig:loadedscatt}}
\end{figure}

Let us consider the case of an antenna located within the diffuse field generated inside an RC (Fig.~\ref{fig:loadedscatt}). In that case, the RC composite $Q$-factor can be expressed as
\begin{equation}
Q_{\mathrm{RC}}^{-1}=Q_{\mathrm{empty}}^{-1}+Q_{\mathrm{a}}^{-1}\label{eq:compositeQ}
\end{equation}
where $Q_{\mathrm{empty}}$ is the chamber $Q$-factor without the antenna inside, and $Q_{\mathrm{a}}$ is the antenna contribution to the RC $Q$-factor. 
The antenna can be viewed as a loaded scatterer, i.e., an antenna with input impedance $Z_{\mathrm{A}}$ terminated by a load impedance $Z_{\mathrm{L}}$. 
In terms of antenna properties, such a system is characterized by two key parameters: the reflection coefficient $\Gamma_{\mathrm{L}}$ and the antenna radiation efficiency $e_{\mathrm{r}}$. 
On the one hand, the reflection coefficient quantifies the reflection that occurs between the antenna and the load and is given by $\Gamma_{\mathrm{L}}=(Z_{\mathrm{L}}-Z_{\mathrm{A}}^{*})/(Z_{\mathrm{L}}+Z_{\mathrm{A}})$ where the superscript $^{*}$ denotes the complex conjugate. 
On the other hand, the antenna radiation efficiency, denoted here $e_{\mathrm{r}},$ is defined (in the transmitting case) as the ratio of the total power radiated by the antenna to the net power accepted from a transmitter with impedance $Z_{\mathrm{L}}$ connected to the antenna port \cite{IEEE2014}.
Consequently, $e_{\mathrm{r}}$ is independent of $\Gamma_{\mathrm{L}}$. 

Based on these two antenna characteristics, an initial model for the antenna's contribution to the RC $Q$-factor was derived by D.~Hill in \cite{Hill1998} as follows:
\begin{equation}
\textrm{{D.\,Hill\,[9]\,\,\,} }\frac{Q_{0}}{Q_{\mathrm{a,Hill}}}=e_{\mathrm{r}}(1-\left|\Gamma_{\mathrm{L}}\right|^{2})\label{eq:Qa_hill}
\end{equation}
where $Q_{0}=16\pi^{2}V/\lambda^{3}$ is the antenna's contribution to the RC $Q$-factor in the case of a perfectly-matched and ideally-efficient (lossless) antenna, $V$ is the RC volume and $\lambda$ is the wavelength. This model was developed to evaluate the power dissipated in the load of a receiving antenna \cite{Hill1998} and has been widely used in the literature. 
It is based on the premise that a portion of the power impinging on the antenna reaches the load and is dissipated. 
Indeed, according to the loss mechanism labelled \raisebox{.5pt}{\textcircled{\raisebox{-.9pt} {2}}} in Fig.~\ref{fig:loadedscatt}, the power dissipated by the load is equal to $\left<P_{\mathrm{r}}\right>e_{\mathrm{r}}(1-\left|\Gamma{}_{\mathrm{L}}\right|^{2})$ where $\left<P_{\mathrm{r}}\right>$ is the average power received by the antenna within the RC. 
As illustrated in Fig.~\ref{fig:loadedscatt} and highlighted in \cite{Cozza2018}, this model does not fully account all loss mechanisms. 
For instance, a poorly-efficient antenna ($e_{\mathrm{r}}\longrightarrow0$) should, in practice, induce significant losses within the chamber but, according to (\ref{eq:Qa_hill}) \cite{Hill1998}, it would be considered transparent. 
Indeed, when $e_{\mathrm{r}}\rightarrow0$, the model predicts that $Q_{\mathrm{a,Hill}}\rightarrow\infty$. 
To address this limitation, a refined definition was introduced in 2018 \cite{Cozza2018} incorporating three dissipation processes illustrated in Fig.~\ref{fig:loadedscatt}. First, part of the average received power is dissipated because of the antenna's radiation efficiency (event \raisebox{.5pt}{\textcircled{\raisebox{-.9pt} {1}}}). 
Then, some of the power received at the antenna load is dissipated, depending on $\Gamma_{\mathrm{L}}$ (event \raisebox{.5pt}{\textcircled{\raisebox{-.9pt} {2}}}).
Finally, the power reflected by the load is transmitted back to the RC, through some losses due, again, to the radiation efficiency (event \raisebox{.5pt}{\textcircled{\raisebox{-.9pt} {3}}}). 
This refinement led to the new formulation:
\begin{equation}
\textrm{{A.\,Cozza\,[18]\,\,}}\frac{Q_{0}}{Q_{\mathrm{a,Cozza}}}=1-e_{\mathrm{r}}^{2}\left|\Gamma_{\mathrm{L}}\right|^{2}.\label{eq:Qa_Cozza}
\end{equation}
Taking again the case of a poorly-efficient antenna ($e_{\mathrm{r}}\rightarrow 0$), the model in (\ref{eq:Qa_Cozza}) does account for these losses, as $Q_{\mathrm{a,Cozza}}\rightarrow1$.
This refined model has been indirectly validated by experimentally showing that the estimated radiation efficiency does not depend on the load impedance \cite{Cozza2018}.

The two aforementioned models rely on a power balance approach that treats losses during reception, reflection on the load, and re-emission as independent processes. 
Therefore, they completely neglect losses associated with the antenna structure, i.e., losses which do not depend on $\Gamma_{\mathrm{L}}$ and $e_{\mathrm{r}}$. 
However, according to antenna scattering theory and related works on antenna RCS \cite{Knott2004}, the antenna structure has an impact on the overall losses brought by the antenna. 
To illustrate this idea, consider a horn antenna with a piece of absorbing material attached to the outside of the waveguide.
Such an absorber would not affect the antenna's radiation efficiency or its reflection coefficient; yet, it would significantly increase the losses brought by such an antenna within an RC. 
As suggested in \cite{Hill1994} (Section~V), some additional losses can be considered to compensate for the scattering losses brought by antennas.
However, such an approach fails to capture potential interference phenomena between the antenna and the structural contributions, which could significantly alter the overall power budget. 
Indeed, because of these interference effects, a dipole antenna still strongly interacts with the incident field when short-circuited ($\Gamma_{\mathrm{L}}=-1$) whereas it is almost transparent when open-circuited ($\Gamma_{\mathrm{L}}=1$).
Such behavior cannot be predicted by the two existing models (which only depend on the modulus squared of $\Gamma_{\mathrm{L}}$) nor by considering the antenna and the structural contributions as independent processes. 
Therefore, in the following sections, we introduce a new derivation to model the antenna contribution to the RC $Q$-factor, accounting for both structural and antenna modes as well as potential interferences between them.

\section{Derivation of the antenna contribution to the RC $Q$-factor}

\label{sec:newmodel}

This section aims at deriving a new model for the antenna contribution
to the RC $Q$-factor. It begins with its definition and its relation
to the antenna AACS. Then, after a brief review of the SWE formalism,
the antenna is described in terms of a scattering matrix where the
ingoing and outgoing waves are expressed as spherical harmonics. From
this representation, the antenna ACS is derived, followed by its spatially-averaged
value. Finally, a new model for the antenna contribution to the RC
$Q$-factor is computed from the AACS, highlighting the existence
of a structural component.

\subsection{Definition of the antenna contribution to the RC $Q$-factor}

If an antenna is placed within an electrically-large cavity such as
an RC, its contribution to the RC $Q$-factor $Q_{\mathrm{a}}$ is
defined as:
\begin{equation}
Q_{\mathrm{a}}=\omega U/P_{\mathrm{d}}
\end{equation}
where $\omega$ is the angular frequency, $U$ is the steady-state
energy in the cavity and $P_{\mathrm{d}}$ is the power dissipated
by the antenna. It has been shown in \cite{Hill1994} that $Q_{\mathrm{a}}$
can be related to the antenna's average absorption cross-section (AACS)
$\bar{\sigma}_{\mathrm{abs}}$ as:
\begin{equation}
\frac{Q_{0}}{Q_{\mathrm{a}}}=\frac{8\pi\bar{\sigma}_{\mathrm{abs}}}{\lambda^{2}}\label{eq:Q_aacs}
\end{equation}
where $\lambda$ is the wavelength. Therefore, the objective of the
following subsections is to derive $\bar{\sigma}_{\mathrm{abs}}$.

\subsection{Vectorial field spherical harmonic decomposition}

\label{subsec:SWE}

This subsection briefly reviews the SWE formalism used throughout
this paper and described in \cite{Hansen1988}. The vectorial electric
field $\mathbf{E}$ at a position $\mathbf{r}$ can be decomposed
over converging $\mathbf{E}_{-}$ and diverging $\mathbf{E}_{+}$
fields expressed in terms of spherical harmonics such as:
\begin{equation}
\mathbf{E}(\mathbf{r})=\underbrace{\sum_{smn}c_{smn}\mathbf{F}_{smn}^{(4)}(\mathbf{r})}_{\mathbf{E}_{-}}+\underbrace{\sum_{smn}d_{smn}\mathbf{F}_{s,m,n}^{(3)}(\mathbf{r})}_{\mathbf{E}_{\mathrm{+}}}\label{eq:E_vec}
\end{equation}
where $\mathbf{F}_{smn}^{(3)}(\mathbf{r})$ and $\mathbf{F}_{smn}^{(4)}(\mathbf{r})$
are the spherical Hankel wave functions of first (diverging) and second
(converging) kind, of degree $n=\{1,2,3,...,N\}$ and of order $m=\{-n,-n+1,...,0,...,n-1,n\}$,
$c_{smn}$ and $d_{smn}$ are the complex coefficients of the wave
functions, $s$ is the index indicating whether it is a TE-wave ($s=1)$
or a TM-wave ($s=2)$ coefficient. Note that $\mathbf{F}_{smn}^{(3)*}(\mathbf{r})=(-1)^{m}\mathbf{F}_{s,-m,n}^{(4)}(\mathbf{r})$
(see Eq. (A1.54) in \cite{Hansen1988}). For sake of simplicity, new
linear indices $i$ and $j$ are introduced which combines the three
indexes $s$, $m$ and $n$ in a univoke manner as $\left\{ i,j\right\} =2\left[n(n+1)+m-1\right]+s$,
so that (\ref{eq:E_vec}) can be written as
\begin{equation}
\mathbf{E}(\mathbf{r})=\underbrace{\sum_{j=0}^{J}c_{j}\mathbf{F}_{j}^{(4)}(\mathbf{r})}_{\mathbf{E}_{-}}+\underbrace{\sum_{i=0}^{J}d_{i}\mathbf{F}_{i}^{(3)}(\mathbf{r})}_{\mathbf{E}_{\mathrm{+}}}\label{eq:E_vec_j}
\end{equation}

\subsection{Antenna scattering matrix}

\begin{figure}
\begin{centering}
\includegraphics[width=0.8\columnwidth]{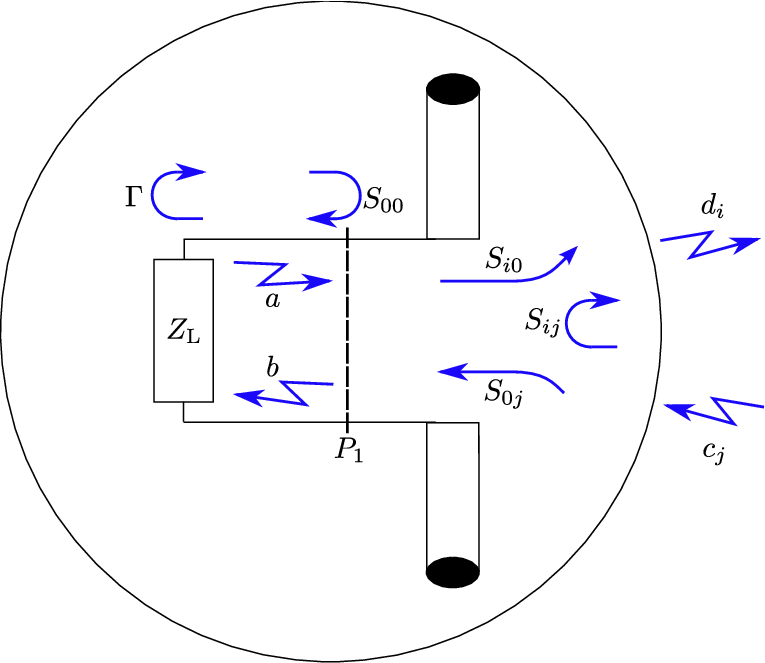}
\par\end{centering}
\caption{Antenna scattering matrix representation.\textcolor{red}{{} }\label{fig:Ssyst}}
\end{figure}

We consider an antenna scattering system, first introduced by R. H.
Dicke in 1947 \cite{Dicke1947} and later taken up by R. E. Collin
in his textbook \cite{Collin1969} whose representation is shown in
Fig.~\ref{fig:Ssyst}. The antenna is fed by a single transmission
line or waveguide that supports only one propagating mode. The incident
wave (towards the terminal plane $P_{1})$ and reflected wave (towards
the load impedance $Z_{\mathrm{L}}$) in the feed line have amplitudes
$a$ and $b$, respectively. Since the antenna is driven by external
waves and a linear load $Z_{\mathrm{L}}$ is attached to the port,
then

\begin{equation}
a=\Gamma b\label{eq:aGammab}
\end{equation}
where $\Gamma$ is the reflection coefficient at the load interface,
as seen from the reference plane. It \lyxdeleted{Francois SARRAZIN}{Mon Apr  7 11:49:36 2025}{ }is
defined with respect to an arbitrary reference impedance $Z_{0}$
as $\Gamma=(Z_{\mathrm{L}}-Z_{0}^{*})/(Z_{\mathrm{L}}+Z_{0})$.

The field outside a sphere of radius $r=r_{0}$, enclosing the antenna,
can be expanded into converging and diverging propagating spherical
modes with complex coefficients $c_{j}$ and $d_{i}$, respectively,
according to the formalism presented in subsection~\ref{subsec:SWE}.
Due to linearity, the input channels are linearly related to the output
ones through the scattering matrix $\mathbf{S}$ defined as:
\begin{equation}
\left[\begin{array}{c}
b\\
d_{1}\\
\vdots\\
d_{J}
\end{array}\right]=\underbrace{\left[\begin{array}{cccc}
S_{00} & S_{01} & \cdots & S_{0J}\\
S_{10} & S_{11} & \cdots & S_{1J}\\
\vdots & \vdots & \ddots & \vdots\\
S_{J0} & S_{J1} & \cdots & S_{JJ}
\end{array}\right]}_{\mathbf{\mathbf{S}}}\left[\begin{array}{c}
a\\
c_{1}\\
\vdots\\
c_{J}
\end{array}\right].\label{eq:scat_mat}
\end{equation}
Please note that, because reciprocity holds, $\mathbf{S}$ is symmetric,
meaning $S_{ij}=S_{ji}$. In the following, we aim to express the
coefficients $c_{j}$ and $d_{i}$ in terms of the scattering parameters.
The approach is similar to previous works on plane-wave scattering-matrix
theory \cite{Kerns1976} or minimum-scattering antennas \cite{Kahn1965}.
Our main contribution will emerge when we average the derived ACS
in Section~\ref{subsec:AACS}.

From (\ref{eq:scat_mat}), $b$ can be expressed as the sum of the
reflected wave\textcolor{black}{s }at the terminal impedance and the
contribution of the external converging waves such as:
\begin{equation}
b=S_{00}a+\sum_{j=1}^{J}S_{0j}c_{j},\label{eq:b}
\end{equation}
where $J$ is the maximum index sum that is roughly equal to $2\left\lfloor kr_{0}\right\rfloor ^{2}$,
$k=2\pi/\lambda$ being the wavenumber. Also, as $a$ and $b$ are
related to each other through (\ref{eq:aGammab}) , (\ref{eq:b})
can be rewritten as
\begin{equation}
b=\frac{1}{(1-S_{00}\Gamma)}\sum_{j=1}^{J}S_{0j}c_{j}.\label{eq:b_2}
\end{equation}
Following the same process, $d_{i}$ can be expressed from (\ref{eq:scat_mat}) as:
\begin{equation}
d_{i}=S_{i0}\Gamma b+\sum_{j=1}^{J}S_{ij}c_{j}.
\end{equation}
Replacing $b$ by its expression from (\ref{eq:b_2}) yields:
\begin{equation}
d_{i}=\frac{S_{i0}\Gamma}{(1-S_{00}\Gamma)}\sum_{j=1}^{J}S_{0j}c_{j}+\sum_{j=1}^{J}S_{ij}c_{j}.
\end{equation}
Please note that $S_{00}$ is the antenna reflection coefficient defined with respect to the reference impedance $Z_{0}$ as $S_{00}=(Z_{\mathrm{A}}-Z_{0}^{*})/(Z_{\mathrm{A}}+Z_{0})$.
For simplicity, but without loss of generality, we assume $Z_{0}=Z_{\mathrm{A}}$, which gives $S_{00}=0$. Additionnally, we can define $\Gamma_{\mathrm{L}}$ as equivalent to $\Gamma$ when $Z_{0}=Z_{\mathrm{A}}$, i.e., $\Gamma_{\mathrm{L}}=(Z_{\mathrm{L}}-Z_{\mathrm{A}}^{*})/(Z_{\mathrm{L}}+Z_{\mathrm{A}})$, therefore
\begin{equation}
d_{i}=S_{i0}\Gamma_{\mathrm{L}}\sum_{j=1}^{J}S_{0j}c_{j}+\sum_{j=1}^{J}S_{ij}c_{j}=\sum_{j=1}^{J}S_{ij}^{\prime}c_{j}\label{eq:d_S}
\end{equation}
where
\begin{equation}
S_{ij}^{\prime}=S_{i0}\Gamma_{\mathrm{L}}S_{0j}+S_{ij}.\label{eq:Sp}
\end{equation}

The primed scattering parameters can be viewed as an extension of the unprimed scattering parameters in which the portion of energy reflected back by the load impedance is taken into account. Please note that the $S_{ij}^{\prime}$ are defined only for $\left\{ i,j\right\} \geq1$ and assuming that $Z_{0}=Z_{\mathrm{A}}$.

\subsection{Antenna absorption cross-section}

The ACS $\sigma_{\mathrm{abs}}^{p}(\Omega_{0})$ of any object, illuminated by a plane wave of polarization $p$ arriving from the incident solid angle $\Omega_{0}$, is defined as the ratio between the power absorbed by this object $P_{\mathrm{abs}}$ and the intensity $I_{0}$ of the incident plane wave. 
The absorbed power is deduced here from the difference between the ingoing power $P_{\mathrm{in}}$ and the outgoing power
$P_{\mathrm{out}}$ so that
\begin{equation}
\sigma_{\mathrm{abs}}^{p}(\Omega_{0})=\frac{P_{\mathrm{abs}}}{I_{0}}=\frac{P_{\mathrm{in}}-P_{\mathrm{out}}}{I_{0}}.\label{eq:sig_a}
\end{equation}

The ingoing power can be computed as the sum of all the converging field components, $P_{\mathrm{in}}$ is therefore given by
\begin{equation}
P_{\mathrm{in}}=\frac{1}{2\eta}\oint_{S}\left\Vert E_{-}(\mathbf{r})\right\Vert ^{2}\mathrm{d}^{2}r,
\end{equation}
where $\eta$ is the wave impedance and the integral is performed
over a close surface $S$ surrounding the antenna. Assuming that the far-field condition is fulfilled, the ingoing power can be expressed in a very simple way as (see Section 2.2.4 in \cite{Hansen1988})
\begin{equation}
P_{\mathrm{in}}=\frac{1}{2\eta k^{2}}\sum_{j=1}^{J}\left|c_{j}\right|^{2}.\label{eq:Pin}
\end{equation}
 In the case of a plane wave, the expansion coefficients $c_{j}$
are known to be equal to (see Section~A1.6 in \cite{Hansen1988})
\begin{equation}
c_{j}=c_{smn}=\frac{1}{2}\sum_{smn}(-1)^{m}\sqrt{4\pi}\,\mathrm{i}\,\mathbf{E}_{0}\cdot\mathbf{K}_{s,-m,n}(\Omega_{0})\label{eq:cj}
\end{equation}
where $\mathbf{E}_{0}$ is the incident electric field and $\mathbf{K}_{s,-m,n}$ are the far-field pattern functions. Please note that the expansion coefficients depend on $\Omega_{0}$ . Following the same steps as for $P_{\mathrm{in}}$, the power $P_{\mathrm{out}}$ is given by
\begin{equation}
P_{\mathrm{out}}=\frac{1}{2\eta}\oint_{S}\left\Vert E_{\mathrm{div}}(\mathbf{r})\right\Vert ^{2}\mathrm{d}^{2}r=\frac{1}{2\eta k^{2}}\sum_{j=1}^{J}\left|d_{j}\right|^{2}.\label{eq:Pout}
\end{equation}

Noticing that the incident intensity can be expressed as $I_{0}=\left\Vert \mathbf{E}_{0}\right\Vert /2\eta$,
injecting (\ref{eq:Pin}) and (\ref{eq:Pout}) into (\ref{eq:sig_a}), and replacing $d_{j}$ by its expression from (\ref{eq:d_S}) yields
\begin{equation}
\sigma_{\mathrm{abs}}^{p}(\Omega_{0})=\frac{1}{k^{2}\left\Vert \mathbf{E}_{0}\right\Vert }\sum_{i=1}^{J}\left(\left|c_{i}\right|^{2}-\left|\sum_{j=1}^{J}S_{ij}^{\prime}c_{j}\right|^{2}\right).
\end{equation}

\subsection{Average absorption cross-section}

\label{subsec:AACS}

The average absorption cross-section (AACS) $\bar{\sigma}_{\mathrm{abs}}$
is computed from the averaging of $\sigma_{\mathrm{abs}}^{p}(\Omega_{0})$
over the 2 polarizations and the incident solid angle $\Omega_{0}$,
i.e.,
\begin{equation}
\bar{\sigma}_{\mathrm{abs}}=\frac{1}{4\pi}\frac{1}{2}\sum_{p=1}^{2}\int\sigma_{\mathrm{abs}}^{p}(\Omega_{0})\mathrm{d}\Omega_{0}.
\end{equation}

Because of the orthogonality of the far-field pattern functions (as
demonstrated in Appendix~A),
\begin{eqnarray}
\sum_{p=1}^{2}\int c_{j}(\Omega)c_{j'}^{*}(\Omega)\mathrm{d}\Omega & = & 4\pi^{2}\left\Vert \mathbf{E}_{0}\right\Vert ^{2}\delta_{jj'},
\end{eqnarray}
where $\delta_{jj'}$ is the Kronecker delta. As a consequence, the antenna AACS is given by
\begin{equation}
\bar{\sigma}_{\mathrm{abs}}=\frac{\pi}{2k^{2}}\sum_{i=1}^{J}\left(1-\sum_{j=1}^{J}\left|S_{ij}^{\prime}\right|^{2}\right).\label{eq:sig_a_mean}
\end{equation}

\subsection{Antenna contribution to the RC $Q$-factor including a structural
mode}

The antenna contribution $Q_{\mathrm{{a}}}$ to the RC $Q$-factor can be determined from (\ref{eq:Q_aacs}) and (\ref{eq:sig_a_mean}) so that
\begin{equation}
\frac{Q_{0}}{Q_{\mathrm{a}}}=\frac{8\pi\bar{\sigma}_{\mathrm{abs}}}{\lambda^{2}}=\sum_{i=1}^{J}\left(1-\sum_{j=1}^{J}\left|S_{ij}^{\prime}\right|^{2}\right).
\end{equation}

Replacing $S_{ij}^{\prime}$ by its expression from (\ref{eq:Sp}),
and noticing that the antenna radiation efficiency $e_{\mathrm{r}}=\sum_{i}\left|S_{i0}\right|^{2}=\sum_{j}\left|S_{0j}\right|^{2}$, the previous equation can be rewritten:
\begin{eqnarray}
\frac{Q_{0}}{Q_{\mathrm{a}}} & = & \sum_{i=1}^{J}\left(1-\sum_{j=1}^{J}\left|S_{ij}\right|^{2}\right)\nonumber \\
 &  & -e_{\mathrm{r}}^{2}\left|\Gamma_{\mathrm{L}}\right|^{2}-2\sum_{i=1}^{J}\sum_{j=1}^{J}\Re\left(S_{i0}\Gamma_{\mathrm{L}}S_{0j}S_{ij}^{*}\right).\label{eq:Q}
\end{eqnarray}

By definition, and according to the convention introduced in \cite{Green1963}, the structural mode of the antenna's contribution to the RC $Q$-factor corresponds to the antenna $Q$-factor when $\Gamma_{\mathrm{L}}=0$. Accordingly, we introduce the structural component $Q_{\mathrm{s}}$ given by
\begin{equation}
\frac{Q_{0}}{Q_{\mathrm{s}}}\equiv\sum_{i=1}^{J}\left(1-\sum_{j=1}^{J}\left|S_{ij}\right|^{2}\right).\label{eq:Qstr}
\end{equation}

Equation (\ref{eq:Q}) can therefore be rewritten as
\begin{equation}
\frac{Q_{0}}{Q_{\mathrm{a}}}=\underset{\mathrm{structural}}{\underbrace{\frac{Q_{0}}{Q_{\mathrm{s}}}}}-\underset{\mathrm{antenna}}{\underbrace{e_{\mathrm{r}}^{2}\left|\Gamma_{\mathrm{L}}\right|^{2}}}-\underset{\mathrm{interference}}{\underbrace{2\sum_{i=1}^{J}\sum_{j=1}^{J}\Re\left(S_{ij}^{\mathrm{L}}S_{ij}^{*}\right)}},\label{eq:Q_final}
\end{equation}
where $S_{ij}^{\mathrm{L}}=S_{i0}\Gamma_{\mathrm{L}}S_{0j}$ is the scattering matrix that includes only the interaction of the modes with the load impedance. Therefore, it is demonstrated here that the antenna contribution to the RC $Q$-factor consists of three distinct parts:
\begin{itemize}
\item \textbf{Structural:} the first one is the contribution from the antenna structure, which dissipates energy independently of the antenna radiation and impedance properties, i.e., the structural mode,
\item \textbf{Antenna:} the second one is solely due to the antenna's radiation and impedance properties, i.e., the antenna mode,
\item \textbf{Interference:} the third one is the interference between the antenna and structural modes, i.e., the \textit{interference} mode. It is expressed here as a correlation between the scattering component related to the antenna load $S_{ij}^{\mathrm{L}}$ and the scattering component related to the structure $S_{ij}^{*}$.
\end{itemize}
Equation (\ref{eq:Q_final}) constitutes the main result of this paper.

\section{Inherent assumptions of previous models}

\label{sec:comparison}

This section aims at comparing the new model introduced in (\ref{eq:Q_final}) to the two previous models. 
In particular, it focuses on the inherent assumptions that were made. If we consider that the antenna is coupled to only one electromagnetic mode, i.e., $J=1$, therefore (\ref{eq:Q}) can be simplified as
\begin{equation}
\frac{Q_{0}}{Q_{\mathrm{a}}}=\left(1-\left|S_{11}\right|^{2}\right)-e_{\mathrm{r}}^{2}\left|\Gamma_{\mathrm{L}}\right|^{2}-2\Re\left(S_{10}\Gamma{}_{\mathrm{L}}S_{01}S_{11}^{*}\right).
\end{equation}

Note that this mode does not necessarily need to be a harmonic spherical mode but it can be any mode that can be obtained by a unitary transformation.
Then, if we also consider that the antenna does not scatter any energy within the RC, i.e., if we assume that $S_{11}=0$, then
\begin{equation}
\frac{Q_{0}}{Q_{\mathrm{a}}}=1-e_{\mathrm{r}}^{2}\left|\Gamma_{\mathrm{L}}\right|^{2},
\end{equation}
which turns out to be exactly the formula from (\ref{eq:Qa_Cozza})
\cite{Cozza2018}. Therefore, the model from \cite{Cozza2018} is
only valid if we consider that the antenna is coupled to one electromagnetic mode ($J=1$) and if the antenna is considered as a non-scattering object ($Q_{0}/Q_{\mathrm{s}}=1$).

As stated in section~\ref{sec:HillCozza} and in \cite{Cozza2018},
Hill's formulation is inconsistent because it does not consider the radiation efficiency as a loss mechanism, treating it solely as a form of aperture efficiency. In other words, it
assumes that what is captured by the antenna but not transmitted to the load is not dissipated by the antenna. This formulation implicitly assumes that the radiation efficiency is equal to 100~\%. 
Therefore, if we consider the specific case of $e_{\mathrm{r}}=1$, the two models from \cite{Hill1998} and \cite{Cozza2018} lead to the same result:-
\begin{equation}
\frac{Q_{0}}{Q_{\mathrm{a}}}=1-\left|\Gamma_{\mathrm{L}}\right|^{2}.
\end{equation}

\section{Numerical validation}

\label{sec:Validation}

This section is dedicated to the validation of the proposed model
through numerical simulations. First, the home-made simulation code
is briefly introduced and verified by analyzing the absorption and
scattering properties of a half-wavelength dipole antenna. Then, the
accuracy of the proposed model is evaluated and compared with the
two existing models.

\subsection{AACS and ASCS of a dipole antenna}

\begin{figure}
\begin{centering}
\includegraphics[width=1\columnwidth]{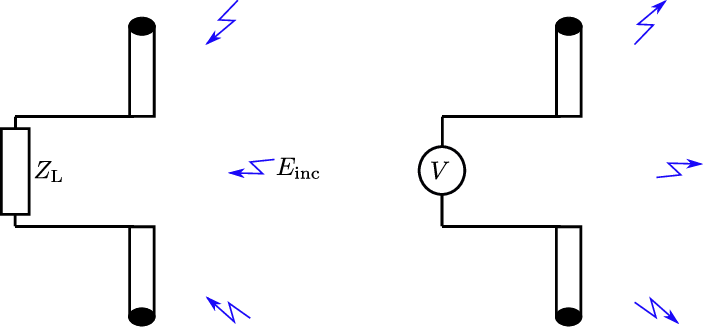}
\par\end{centering}
\caption{Dipole antenna in the receiving mode (left), and in the transmitting
mode (right).\label{fig:Rx_Tx_dipole}}
\end{figure}

A Python implementation based on the Method of Moments (MoM) described in Orfanidis's online book \cite{Orfanidis2016}, together with the associated MATLAB toolbox, has been developed to validate the proposed model through numerical results.
We consider a dipole antenna of length 0.48~$\lambda$ and diameter $5\times10^{-4},\lambda$ operating at 300~MHz. The antenna is discretized into 149 segments, and a resistance per unit length $R_{\Omega}$ is introduced to account for losses.
For simulations in the transmitting mode (Fig.~\ref{fig:Rx_Tx_dipole}~(right)), the antenna is excited by a delta-gap voltage source. The radiation efficiency and reflection coefficient are obtained numerically by solving Hall\'{e}n's equation.
For simulations in the receiving mode, the dipole antenna is terminated with a load impedance (Fig.~\ref{fig:Rx_Tx_dipole}~(left)) and illuminated by an incident plane wave with polarization $\mathbf{E}_{\mathrm{inc}}$. The scattered far field $\hat{\mathbf{E}}_{\mathrm{s}}(\Omega)$ and the current distribution are computed via numerical inversion of Pocklington's equation.

The ACS, $\sigma_{\mathrm{abs}}$ is computed using (\ref{eq:sig_a}) where the power dissipated by ohmic losses is deduced from the current distribution, the real part of the load impedance and $R_{\Omega}$.
To validate the accuracy of the simulation, we also compute two additional cross-sections, the scattering ($\sigma_{\mathrm{sca}}$) and extinction ($\sigma_{\mathrm{ext}}$) cross-sections. These are respectively compited from \cite{Berg2008}:
\begin{eqnarray}
\sigma_{\mathrm{sca}} & = & \frac{\int_{4\pi}\left\Vert \hat{\mathbf{E}}_{\mathrm{sca}}(\Omega)\right\Vert ^{2}d\Omega}{\left\Vert \mathbf{E}_{\mathrm{inc}}\right\Vert ^{2}},\\
\sigma_{\mathrm{ext}} & = & \frac{4\pi\Im\left(\mathbf{E}_{\mathrm{inc}}^{*}\cdot\hat{\mathbf{E}}_{s}(\Omega=\Omega_{\mathrm{in}c})\right)}{\left\Vert \mathbf{E}_{\mathrm{inc}}\right\Vert ^{2}k}.
\end{eqnarray}

Due to the circular symmetry and the fact that the dipole interacts
only with an incident field that is, at least partially, vertically
polarized, it is sufficient to integrate over the elevation angle
only with a 1-degree step to estimate the averaged absorption ($\bar{\sigma}_{\mathrm{abs}}$),
scattering ($\bar{\sigma}_{\mathrm{sca}}$) and extinction ($\bar{\sigma}_{\mathrm{ext}}$)
cross-sections, respectively.

\begin{figure}
\begin{centering}
\includegraphics[width=1\columnwidth]{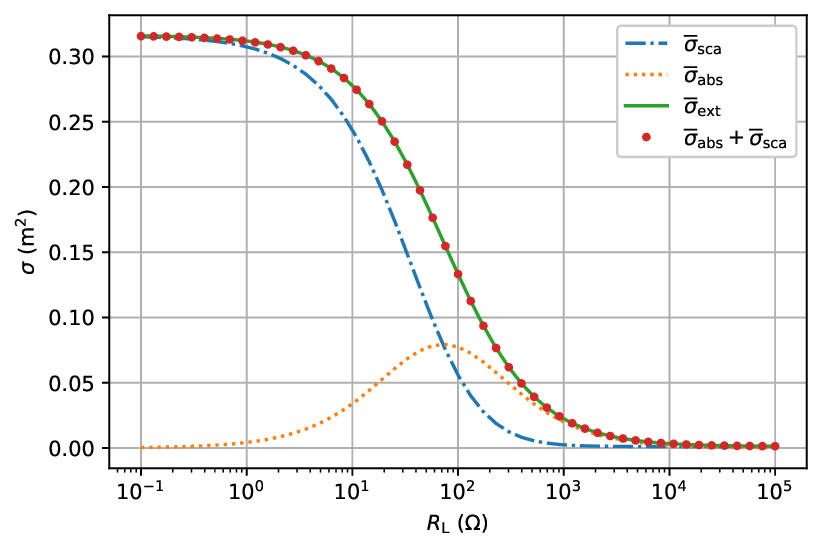}
\par\end{centering}
\begin{centering}
\includegraphics[width=1\columnwidth]{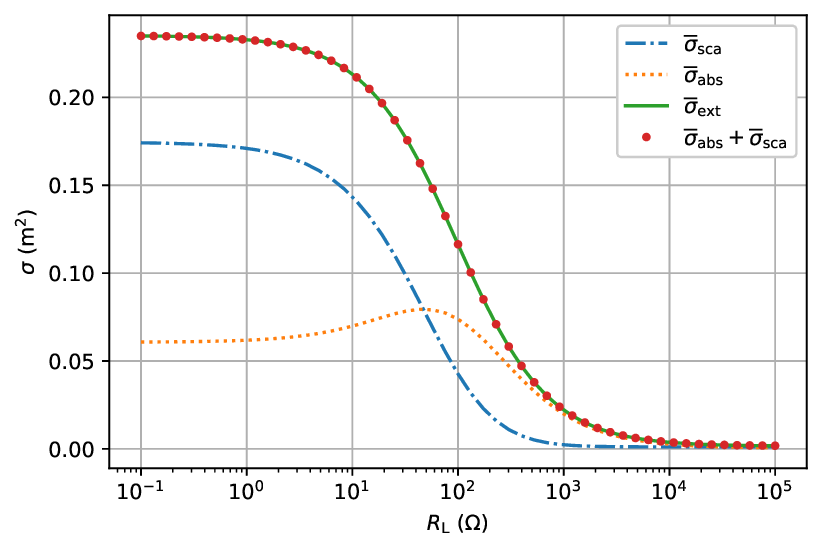}
\par\end{centering}
\begin{centering}
\includegraphics[width=1\columnwidth]{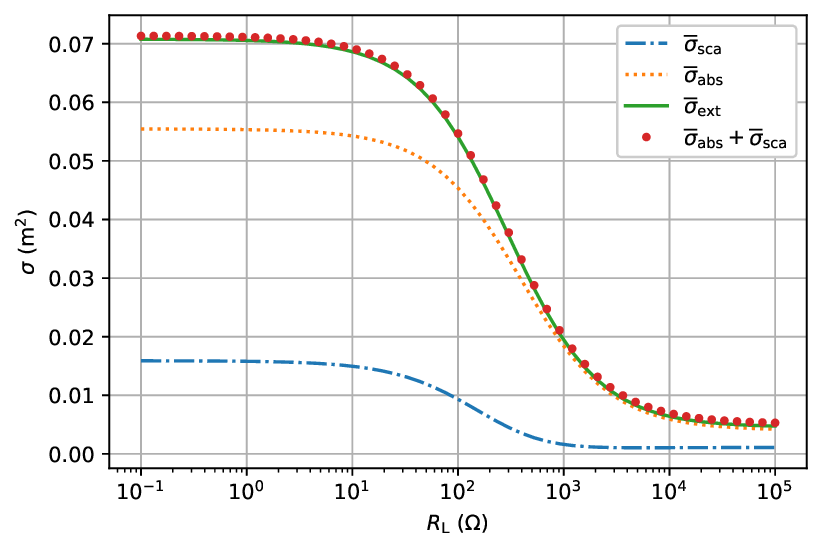}
\par\end{centering}
\centering{}\caption{Average scattering, absorption and extinction cross-sections as a
function of $R_{\mathrm{L}}$ for three cases: (top) $R_{\Omega}=0$~$\Omega\mathrm{m}^{-1}$
($e_{\mathrm{r}}=1$), (center) $R_{\Omega}=100$~$\Omega\mathrm{m}^{-1}$
($e_{\mathrm{r}}=0.75$) and (bottom) $R_{\Omega}=1000$~$\Omega\mathrm{m}^{-1}$
($e_{\mathrm{r}}=0.22$).\label{fig:abs_scat_ext}}
\end{figure}

Fig.~\ref{fig:abs_scat_ext} (top) presents the three different averaged
cross-sections as a function of a purely-real load impedance $R_{\mathrm{L}}$
for an ideally-efficient antenna ($R_{\Omega}=0$~$\Omega\mathrm{m}^{-1}$).
In this case, $e_{\mathrm{r}}=1$ and $Z_{\mathrm{A}}=72.1\,\Omega+\mathrm{i}0.43\,\Omega$.
It is observed that the sum of $\bar{\sigma}_{\mathrm{abs}}$ and
$\bar{\sigma}_{\mathrm{sca}}$ overlaps with $\bar{\sigma}_{\mathrm{ext}}$,
as expected theoretically (see e.g., \cite{Berg2008}) which confirms
that our simulation is fully consistent. Also, $\bar{\sigma}_{\mathrm{abs}}$
and $\bar{\sigma}_{\mathrm{sca}}$ are equal when the antenna is matched
($R_{\mathrm{L}}\sim72\,\Omega$), confirming that under matched condition, the dipole antenna scatters as much energy as it absorbs. 
Finally, the short-circuit and open-circuit cases yield the same level of absorption, which is consistent since mismatch is the only loss mechanism.

Fig.~\ref{fig:abs_scat_ext} (center) presents the same results but for a lossy antenna ($R_{\Omega}=100$~$\Omega\mathrm{m}^{-1}$).
In this case, $e_{\mathrm{r}}=0.75$ and $Z_{\mathrm{A}}=96.9\,\Omega-\mathrm{i}2.75\,\Omega$.
Unlike the previous case, it is observed that $\bar{\sigma}_{\mathrm{abs}}$ of a short-circuited dipole antenna is not zero and is therefore different
from that of the open-circuited dipole antenna. Such behavior could not be modeled by the previous $Q$-factor derivations, as they only depend on the modulus of the reflection coefficient, i.e., $\left|\Gamma_{\mathrm{L}}\right|$.
Finally, Fig.~\ref{fig:abs_scat_ext} (bottom) presents the same
results but for a highly lossy antenna ($R_{\Omega}=1000$~$\Omega\mathrm{m}^{-1}$).
In this case, $e_{\mathrm{r}}=0.22$ and $Z_{\mathrm{A}}=304\,\Omega-\mathrm{i}65.8\,\Omega$.
Here, it is seen that $\bar{\sigma}_{\mathrm{{ext}}}$ is much lower than in the two previous cases, and the absorption dominates, regardless of the load impedance. 
Please note that the loss resistance is the only source of losses within the system in our MoM-based simulation.
Therefore, its value has been chosen solely to target a range of radiation efficiencies (100\%, 75\% and 22\%) and does not intend to reflect true material conductivity.

\subsection{Comparison of the three models}

This subsection is dedicated to evaluating the accuracy of the introduced antenna $Q$-factor model using MoM numerical results and comparing it with the two previous models. 
For the former models from \cite{Hill1998} and \cite{Cozza2018}, only three parameters are required: the antenna radiation efficiency $e_{\mathrm{r}}$, and the real and imaginary parts of the reflection coefficient $\Gamma_{\mathrm{L}}$, which can be computed from simulations in the transmitting mode. 

Regarding the introduced model from (\ref{eq:Q_final}), it can be
rewritten here as follows:
\begin{equation}
\frac{Q_{0}}{Q_{\mathrm{a}}}=\frac{Q_{0}}{Q_{\mathrm{s}}}-e_{\mathrm{r}}^{2}\left|\Gamma_{\mathrm{L}}\right|^{2}-2\left(\Re\left(\Gamma_{\mathrm{L}}\right)\Re\left(C\right)-\Im\left(\Gamma_{\mathrm{L}}\right)\Im\left(C\right)\right)
\end{equation}
where
\begin{equation}
C=\sum_{i=1}^{J}\sum_{j=1}^{J}S_{i0}S_{0j}S_{ij}^{*}.\label{eq:C}
\end{equation}
Therefore, one needs to estimate three more parameters: $Q_{0}/Q_{\mathrm{s}}$, $\Re\left(C\right)$ and $\Im\left(C\right)$. These can be obtained from simulations in the receiving mode, by computing $Q_{0}/Q_{\mathrm{a}}$ for three different load conditions. If $Z_{\mathrm{L}}=Z_{\mathrm{A}}^{*}$, then $\Gamma_{\mathrm{L}}=0$ and therefore
\begin{equation}
\frac{Q_{0}}{Q_{\mathrm{s}}}=\frac{Q_{0}}{Q_{\mathrm{a,\Gamma_{\mathrm{L}}=0}}}.\label{eq:Qstrcut_MoM}
\end{equation}
If $Z_{\mathrm{L}}\rightarrow\infty$, then $\Gamma_{\mathrm{L}}=1$, and therefore $\Re\left(C\right)$ can be evaluated through
\begin{equation}
\Re\left(C\right)=-\frac{1}{2}\left[\frac{Q_{0}}{Q_{\mathrm{a,\Gamma_{\mathrm{L}}=1}}}-\frac{Q_{0}}{Q_{\mathrm{s}}}+e_{\mathrm{r}}^{2}\right].
\end{equation}
Finally, if $Z_{\mathrm{L}}=(Z_{\mathrm{A}}^{*}+\mathrm{iZ_{A}})/(1-\mathrm{i})$, then $\Gamma_{\mathrm{L}}=\mathrm{i}$, and $\Im\left(C\right)$ can be estimated as
\begin{equation}
\Im\left(C\right)=\frac{1}{2}\left[\frac{Q_{0}}{Q_{\mathrm{a,\Gamma_{\mathrm{L}}^{'}=\mathrm{i}}}}-\frac{Q_{0}}{Q_{\mathrm{s}}}+e_{\mathrm{r}}^{2}\right].
\end{equation}

\begin{figure}
\begin{centering}
\includegraphics[width=1\columnwidth]{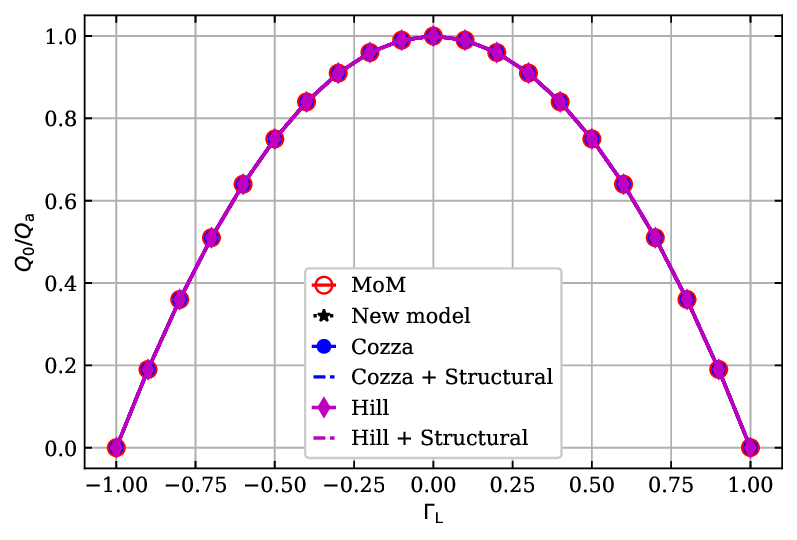}
\par\end{centering}
\begin{centering}
\includegraphics[width=1\columnwidth]{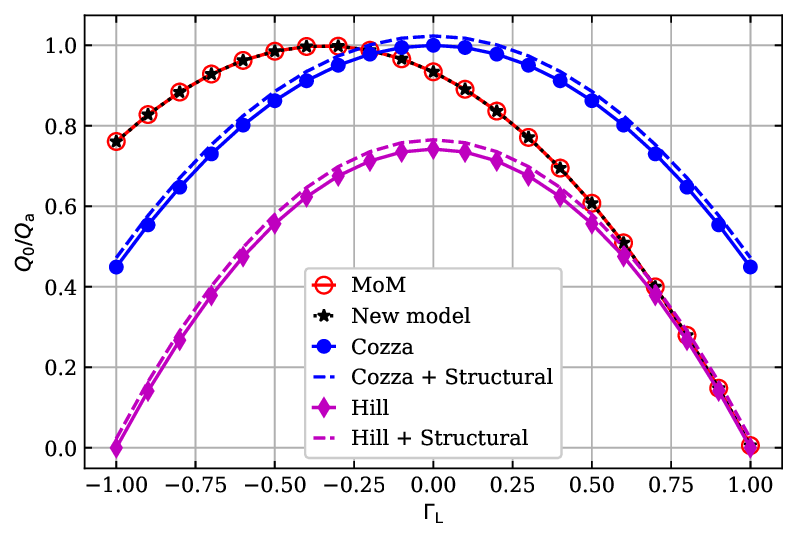}
\par\end{centering}
\begin{centering}
\includegraphics[width=1\columnwidth]{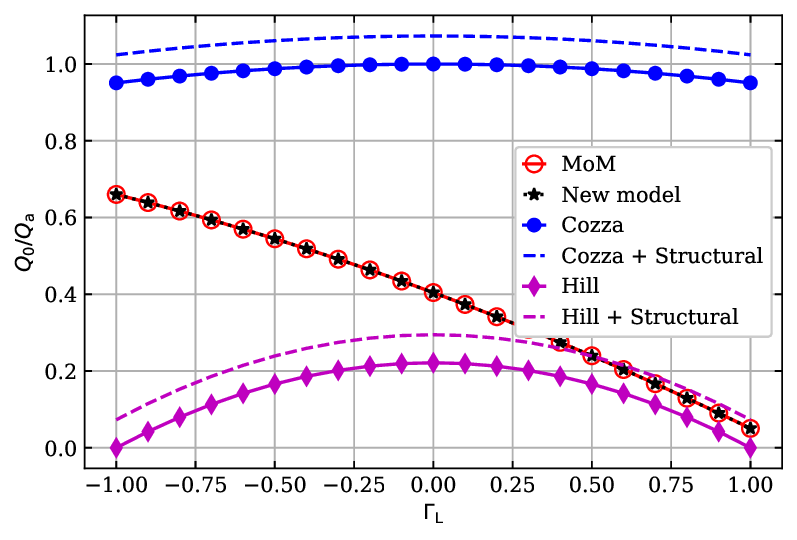}
\par\end{centering}
\caption{Antenna contribution to the RC $Q$-factor as a function of the reflection coefficient for three cases: (top) $R_{\Omega}=0$~$\Omega\mathrm{m}^{-1}$
($e_{\mathrm{r}}=1$), (center) $R_{\Omega}=100$~$\Omega\mathrm{m}^{-1}$
($e_{\mathrm{r}}=0.75$) and (bottom) $R_{\Omega}=1000$~$\Omega\mathrm{m}^{-1}$
($e_{\mathrm{r}}=0.22$).}
\label{fig:Antenna-Q-factor}
\end{figure}

Fig.~\ref{fig:Antenna-Q-factor} presents the antenna's contribution to the RC $Q$-factor as a function of $\Gamma_{\mathrm{L}}$ for the same three cases discussed previously. The results obtained from the MoM simulation are compared with those predicted by the three different models.
For the lossless case shown in Fig.~\ref{fig:Antenna-Q-factor}~(top), all three models yield identical results and are in agreement with the MoM simulation. Indeed, for this particular lossless case, neither the structural mode nor the radiation efficiency introduces losses (a mathematical proof is provided in Appendix~B).
For the two other lossy cases, a very good agreement is observed between the MoM results and the proposed model over the entire range of $\Gamma_{\mathrm{L}}$. In contrast, both existing models produce highly inaccurate results. In particular, they fail to capture the asymmetric behavior observed as a function of $\Gamma_{\mathrm{L}}$.

However, it may be argued that the models proposed by D.~Hill and A.~Cozza were not originally intended to account for structural losses. Indeed, in Section~V of \cite{Hill1994}, it is stated that structural losses introduced by antennas within an RC can be compensated by increasing the wall surface area while reducing the cavity volume according to the antenna size; in other words, by considering the addition of a non-radiating lossy object with dimensions identical to those of the antenna.
Following this approach, we computed the losses that the antenna would introduce if it were replaced by a simple object--here, a metallic cylinder. This methodology is consistent with that commonly used to evaluate the contribution of a mode stirrer or the RC walls to the RC $Q$-factor.
The contribution of an object to the RC $Q$-factor is related to its AACS $\bar{\sigma}_{\mathrm{abs}}^{\mathrm{obj}}$ such as:
\begin{equation}
Q_{\mathrm{obj}}=\frac{2\pi V}{\lambda}\frac{1}{\bar{\sigma}_{\mathrm{abs}}^{\mathrm{obj}}},\label{eq:Q_obj}
\end{equation}
where $V$ is the RC volume, and $\lambda$ the wavelength. Assuming that this object contributes in the same manner as the RC walls, its contribution to the RC $Q$-factor can be expressed as:
\begin{equation}
Q_{\mathrm{obj}}=\frac{3V}{2\mu_{\mathrm{r}}S\delta_{\mathrm{w}}}\label{eq:Q_wall}
\end{equation}
where $\mu_{\mathrm{r}}$ is the relative permeability of the object, $S$ its total surface area, and  $\delta_{\mathrm{w}}$ the skin depth given by:
\begin{equation}
\delta_{\mathrm{w}}=\sqrt{\frac{2R_{\Omega}}{\omega_{0}\mu_{\mathrm{r}}\mu_{0}}}\label{eq:skindepth}
\end{equation}
where $R_{\Omega}$ is the electrical resistivity of the object, and $\mu_{0}$ is the free-space permeability. It follows that:
\begin{equation}
\bar{\sigma}_{\mathrm{abs}}^{\mathrm{obj}}=\frac{4\pi\mu_{\mathrm{r}}S\delta_{\mathrm{w}}}{3\lambda}\label{eq:sigma_abs_obj}
\end{equation}
Such losses were computed for the two aforementioned lossy cases, and their contributions were incorporated to the models of Hill and Cozza in Fig.~\ref{fig:Antenna-Q-factor}. Note that this contribution is, by definition, independent of $\Gamma_\mathrm{L}$.
As expected, it increases the overall estimated losses introduced by the antenna; however, it does not improve the accuracy of the estimation.

\section{Antenna characteristics retrieval}
\label{sec:Antenna-characteristics-retrieva}
\begin{figure}
\centering{}\includegraphics[width=1\columnwidth]{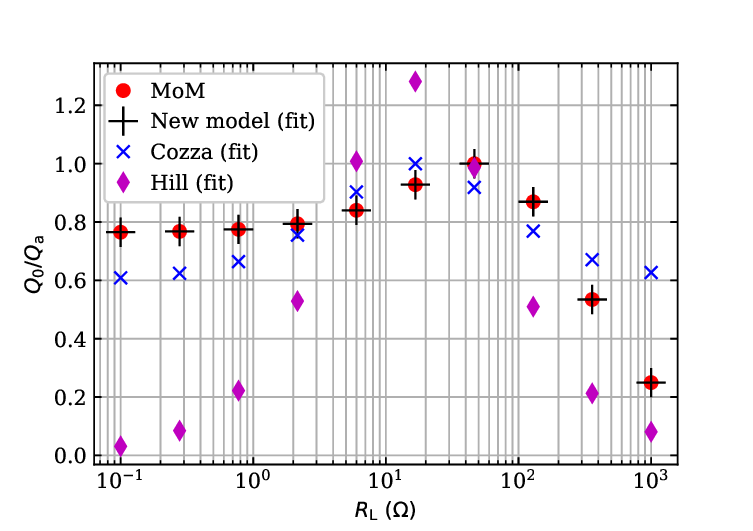}\caption{Estimated antenna contribution to the RC $Q$-factor from a set of 10 purely-real loads $R_{\mathrm{L}}$ ranging from 0.1~$\Omega$ to 1~k$\Omega$.\label{fig:FitReal}}
\end{figure}

This section is dedicated to describing how this model could be used in practice in an RC. 
In particular, we take advantage of the introduced model to retrieve antenna characteristics (i.e., radiation efficiency and input impedance), using multiple RC $Q$-factor evaluations for various $Z_{\mathrm{L}}$. 
This could enable accurate noninvasive antenna characterization \cite{Krouka2022,Galesloot2023}, which is highly relevant for physically and electrically small antenna testing \cite{Icheln1999,Huitema2014,Bories2010}. 
We consider the case of the lossy antenna with $e_{\mathrm{r}}=0.75$ ($R_{\Omega}=100$~$\Omega\mathrm{m}^{-1}$).
A set of $K=10$ numerical simulations in the receiving mode is performed where the load impedance is purely real ranging from 0.1~$\Omega$ to 1~k$\Omega$.
A nonlinear iterative minimization search algorithm was then implemented in order to minimize the function $F_{\mathrm{Model}}$ defined as
\begin{equation}
F_{\mathrm{Model}}=\sum_{k=1}^{K}(\tilde{Q}_{\mathrm{MoM}}^{-1}(Z_{\mathrm{L}}^{k})-\tilde{Q}_{\mathrm{Model}}^{-1}(Z_{\mathrm{L}}^{k}))
\end{equation}
where $Z_{\mathrm{L}}^{k}$ is the $k^{\mathrm{{th}}}$ load impedance, $\tilde{Q}_{\mathrm{MoM}}$ and $\tilde{Q}_{\mathrm{Model}}$ are the estimated antenna contributions to the RC $Q$-factor using the MoM algorithm and the considered model, respectively. 
In the cases of Hill's and Cozza's theories, three parameters need to be determined: the antenna efficiency $e_{\mathrm{r}}$ and the real and imaginary parts of the antenna input impedance $Z_{\mathrm{A}}$.
Three additional parameters are considered in the proposed model: the structural contribution $Q_{0}/Q_{\mathrm{s}}$, as well as the real and imaginary parts of $C$ defined in (\ref{eq:C}). 
Fig.~\ref{fig:FitReal} presents the antenna contribution to the RC $Q$-factor for  purely-real load $R_{\mathrm{L}}$. Results from the MoM simulation are compared with the best fit that has been obtained using the three models. 
It is shown that the new model exhibits a significantly higher accuracy than the two existing models as it closely aligns with the MoM results.

The retrieved antenna characteristics are presented in Table~\ref{tab:ParaReal}. It can be observed that the extracted antenna radiation efficiency and input impedance obtained using the introduced model are consistent with those obtained using the MoM simulation in the transmitting mode, although the relative error regarding $\Im\left(Z_{\mathrm{A}}\right)$ is relatively large (approximately 25~\%). 
As expected, both of the previous models fail to retrieve relevant parameters, even exhibiting nonphysical results such as $e_{\mathrm{r}}>1$.

\begin{table}
~

\caption{Estimated Antenna Parameters From a Set of 10 Purely-Real $Z_{\mathrm{L}}$
Ranging From 0.1~$\Omega$ to 1~k$\Omega$.\label{tab:ParaReal}}

\centering{}%
\begin{tabular}{|c|c|c|c|}
\hline 
  & $e_{\mathrm{r}}^{2}$ & $\Re\left(Z_{\mathrm{A}}\right)$ & $\Im\left(Z_{\mathrm{A}}\right)$\tabularnewline
\hline 
\hline 
MoM  & 0.55 & 96.9 & -2.72\tabularnewline
\hline 
New model & 0.55 & 96.9 & -2.02\tabularnewline
\hline 
Cozza & 0.40 & 17.6 & -0.48\tabularnewline
\hline 
Hill & 1.64 & 16.3 & 0.00\tabularnewline
\hline 
\end{tabular}
\end{table}

\begin{figure}
\begin{centering}
\includegraphics[width=1\columnwidth]{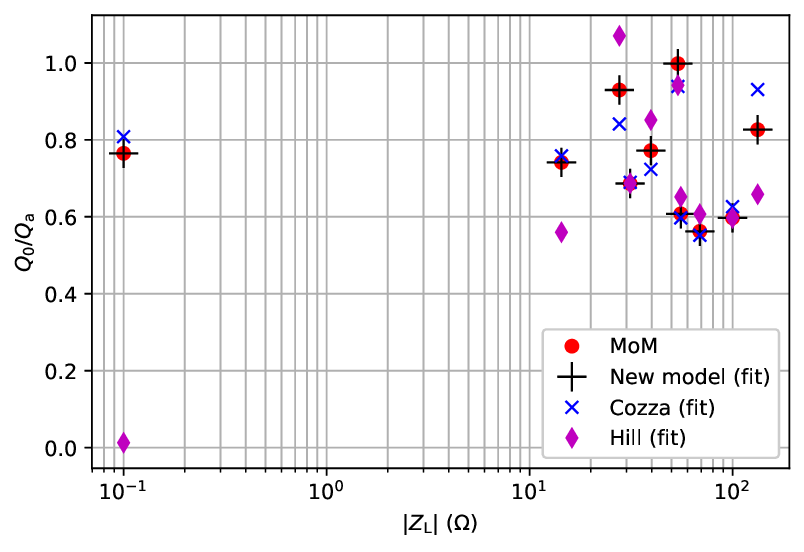}
\par\end{centering}
\caption{Estimated antenna contribution to the RC $Q$-factor from a set of
10 complex-valued $Z_{\mathrm{L}}$.\label{fig:FitComplex}}
\end{figure}

A second study is conducted in which the load impedance $Z_{\mathrm{L}}$ is set to a complex value by adding a transmission line whose length varies linearly from 0~m to 0.4~m between the antenna and a purely real load (still ranging from 0.1~$\Omega$ to 1~k$\Omega$). 
Fig.~\ref{fig:FitComplex} shows the antenna contribution to the RC $Q$-factor obtained from the MoM simulation, compared to the three models after applying the minimization search algorithm. 
Once again, the new model matches the MoM results whereas the two older models lead to inaccurate results.
The relevant extracted parameters are presented in Table~\ref{tab:ParaComplex}.
All parameters are accurately retrieved using the new model. 
In particular, it is demonstrated that $\Im\left(Z_{\mathrm{A}}\right)$ is better estimated than previously thanks to the set of complex-valued $Z_{\mathrm{L}}$.

\begin{table}
\begin{centering}
~\caption{Estimated Antenna Parameters From a Set of 10 Complex-Valued $Z_{\mathrm{L}}$.\label{tab:ParaComplex}}
\par\end{centering}
\centering{}%
\begin{tabular}{|c|c|c|c|}
\hline 
  & $e_{\mathrm{r}}^{2}$ & $\Re\left(Z_{\mathrm{A}}\right)$ & $\Im\left(Z_{\mathrm{A}}\right)$\tabularnewline
\hline 
\hline 
MoM & 0.55 & 96.9 & -2.72\tabularnewline
\hline 
New model& 0.55 & 96.9 & -2.71\tabularnewline
\hline 
Cozza & 0.19 & 117 & -89.4\tabularnewline
\hline 
Hill & 1.18 & 527.5 & 13.2\tabularnewline
\hline 
\end{tabular}
\end{table}

In conclusion, although the improved agreement with the MoM results is not unexpected--given that the proposed model includes three additional parameters--it is shown that the retrieved parameters associated with the antenna characteristics are in very good agreement with those obtained in the transmitting mode. This provides both a further validation of the proposed model and a promising avenue for retrieving antenna characteristics from noninvasive RC measurements.

\section{Validation using full-wave simulations}
\label{sec:fullwave}
\begin{figure}
    \centering
    \includegraphics[width=1.0\columnwidth]{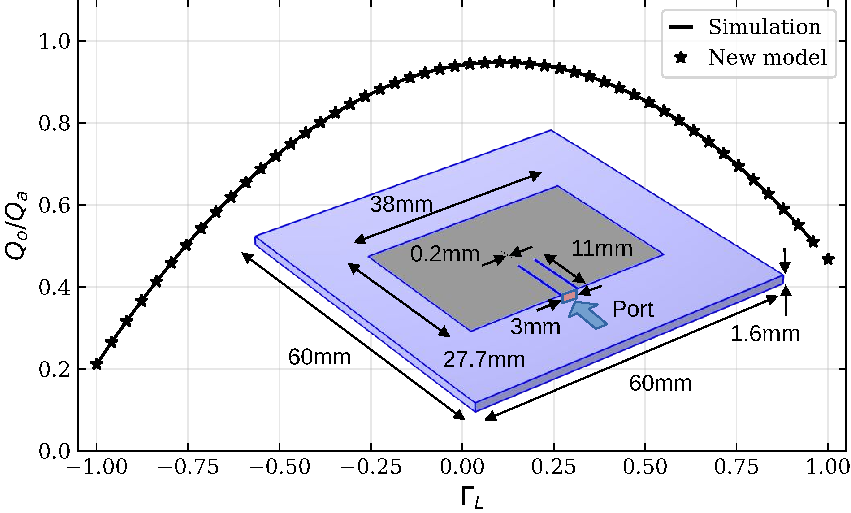}
    \caption{Antenna contribution to the RC $Q$-factor as a function of the reflection coefficient, for a patch antenna simulated using \textit{COMSOL Multiphysics}. The geometry and dimensions of the antenna are shown in the inset. The patch, modeled as a perfect electric conductor (PEC), is placed on an FR4 substrate $(\epsilon = 4.5,\ \tan \delta = 0.03)$. The bottom surface of the substrate is also modeled as a PEC.}
    \label{fig:patch}
\end{figure}

To conclude with a more realistic configuration, full-wave simulations were performed using the Finite Element Method (FEM) implemented in the RF module of \textit{COMSOL Multiphysics} on a patch antenna whose geometry is shown in Fig.~\ref{fig:patch} (inset). The dimensions of the patch were optimized to achieve resonance at 2.45~GHz, while the feeding stub was designed to ensure impedance matching to a 50~$\Omega$ port. The total absorption cross section is computed directly by integrating the Poynting vector over a closed surface in \textit{COMSOL Multiphysics}. The mean absorption cross section over incident directions is then obtained by averaging over 500 directions, uniformly distributed using a geodesic polyhedron, with the directions corresponding to the centers of the triangular facets.

Figure~\ref{fig:patch} shows the antenna contribution to the RC $Q$-factor as a function of the (real) port reflection coefficient, varying from $-1$ to $1$. As in the dipole case, a clear asymmetry is observed, confirming that even when the antenna radiates predominantly over only half of the space, interference effects still significantly influence the antenna contribution to the RC $Q$-factor.
However, several differences can be identified. First, the maximum value reaches approximately $0.95$, compared to nearly $1.0$ for the dipole in the case of $e_\mathrm{r}=0.75$. Moreover, unlike the dipole, the absorption is higher for $\Gamma_L = 1$ than for $\Gamma_L = 0$. This behavior can be explained by the fact that the dimensions of the patch antenna are tuned to achieve resonance when the port is effectively open-circuited (i.e., very high impedance). When the port is instead connected to ground, the antenna resonates at approximately a quarter wavelength, corresponding to a resonance at half the frequency.

Finally, we demonstrate how to passively retrieve the antenna properties from full-wave simulations in the receiving mode. 
To this end, we perform simulations to compute eight values of $Q_{0}/Q_{\mathrm{a}}$ obtained by successively setting the reflection coefficient ($\Gamma_\mathrm{L}$) to $(0.0001, -0.9, 0.9, 0.9j, -0.9j, 0.5, 0.5j, -0.5j$). 
Using these values and applying a minimization algorithm, both the input impedance and the radiation efficiency are accurately recovered: $45~\Omega$ instead of $(47.6 - 2.90j)~\Omega$, and $0.77$ instead of $0.78$ for the radiation efficiency. 
This clearly extends the previous study to significantly more realistic antennas.

\section{Conclusion}

This paper introduced a novel model of the antenna's contribution to the RC $Q$-factor, based on antenna scattering matrix theory.
The model employs spherical wave expansion to account for ingoing and outgoing waves. 
It consists of three components: the structural mode, the antenna mode, and the interference mode. 
The structural mode encompasses all losses arising from the physical structure of the antenna, i.e., not related to its impedance or radiation properties. 
In contrast, the antenna mode accounts for losses solely due to the antenna's characteristics, i.e., radiation efficiency and reflection coefficient. 
Finally, the interference mode is related to the inevitable interaction between the structural and antenna modes. This model is compliant with the conventional decomposition of antenna backscattering, where the structural mode interacts with the antenna mode, leading to either constructive or destructive interference.

The proposed model has been validated through two distinct numerical approaches.
First, simulations based on the Method of Moments were conducted, and the new model was compared with two existing models from the literature (D.~Hill~\cite{Hill1998} and A.~Cozza~\cite{Cozza2018}). These models were found to be inadequate for accurately describing the antenna contribution to the RC $Q$-factor in practical cases, i.e., when $e_{\mathrm{r}} \neq 1$.
Second, full-wave simulations were performed using the Finite Element Method implemented in \textit{COMSOL Multiphysics} on a realistic patch antenna. The results again show very good agreement with the proposed model.

The proposed model has demonstrated its capability to retrieve relevant antenna characteristics from multiple RC $Q$-factor estimations obtained under various antenna loading conditions.
Therefore, it not only enables a more accurate estimation of antenna losses within an RC, but also opens new possibilities for noninvasive antenna characterization.

Future work will include an experimental validation of the proposed model in a reverberation chamber. Such measurements may prove challenging, as they require isolating the antenna contribution to the RC $Q$-factor from the overall composite RC $Q$-factor for potentially small variations of the load impedance. 
Additionally, the model will be extended to the case of multiport antennas.

\section*{Appendices}

\subsection{Orthogonality of the far-field pattern functions}

\label{annexe:OrthogonalityFFfunc}

In this appendix, we demonstrate the orthogonality of the far-field
pattern function defined from (A1.5) in \cite{Hansen1988} as

\begin{eqnarray}
\mathbf{K}{}_{smn}(\theta,\phi) & = & \lim_{kr\rightarrow\infty}\sqrt{4\pi}kre^{-ikr}\text{\ensuremath{\mathbf{F}}}_{smn}^{(3)}(r,\theta,\phi).\label{eq:K}
\end{eqnarray}

To that end, we take the benefit of (A1.69) in \cite{Hansen1988}

\begin{multline}
\oint\left\{ \left(\mathbf{F}_{smn}^{(3)}(r,\theta,\phi).\hat{\boldsymbol{\text{\ensuremath{\theta}}}}\right)\left(\mathbf{F}_{s'm'n'}^{(4)}(r,\theta,\phi).\hat{\boldsymbol{\text{\ensuremath{\theta}}}}\right)\right.\\
\left.+\left(\mathbf{F}_{smn}^{(3)}(r,\theta,\phi).\hat{\text{\ensuremath{\boldsymbol{\phi}}}}\right)\left(\mathbf{F}_{s'm'n'}^{(4)}(r,\theta,\phi).\hat{\text{\ensuremath{\boldsymbol{\phi}}}}\right)\right\} d\Omega\\
=\delta_{ss'}\delta_{m,-m'}\delta_{nn'}(-1)^{m}R_{sn}^{(3)}(kr)R_{sn}^{(4)}(kr)\label{eq:oint}
\end{multline}
where \textbf{$R_{sn}^{(c)}(kr)$} are the radial functions. In the
far-field limit, the expression of the spherical Hankel function of
the first kind can be directly deduced from (\ref{eq:K})

\begin{eqnarray}
\mathbf{F}_{smn}^{(3)}(r,\theta,\phi) & \underset{kr\gg1}{=} & \frac{\mathbf{K}_{smn}}{\sqrt{4\pi}}\frac{e^{ikr}}{kr}
\end{eqnarray}

Moreover, from (A1.14) and (A1.16) \cite{Hansen1988}, it appears
that the far-field expression is given by

\begin{eqnarray}
R_{sn}^{(3)}(kr) & \underset{kr\gg1}{=} & (-i)^{n}(-i)^{2-s}\frac{e^{ikr}}{kr}
\end{eqnarray}

As a consequence,

\begin{eqnarray}
\mathbf{F}_{smn}^{(3)}(r,\theta,\phi) & \underset{kr\gg1}{=} & \frac{R_{sn}^{(3)}(kr)(i)^{n}i^{2-s}}{\sqrt{4\pi}}\mathbf{K}_{smn}(\theta,\phi)\label{eq:F3}
\end{eqnarray}

and

\begin{multline}
\mathbf{F}_{smn}^{(4)}(r,\theta,\phi)\underset{kr\gg1}{=}\\
(-1)^{m}\frac{R_{sn}^{*(3)}(kr)(i)^{-n}i^{s-2}}{\sqrt{4\pi}}\mathbf{K}_{smn}^{*}(\theta,\phi)\label{eq:F4}
\end{multline}

Because $\mathbf{F}_{smn}^{*(3)}(r,\theta,\phi)=(-1)^{m}\mathbf{F}_{s'-m'n'}^{(4)}(r,\theta,\phi)$
(see (A1.54) in \cite{Hansen1988}),

\begin{multline}
\oiint\mathbf{F}_{smn}^{(3)}(r,\theta,\phi).\mathbf{F}_{s'm'n'}^{*(3)}(r,\theta,\phi)=\\
\oiint(-1)^{m}\mathbf{F}_{smn}^{(3)}(r,\theta,\phi).\mathbf{F}_{s'-m'n'}^{(4)}(r,\theta,\phi).\label{eq:F3F3F3F4}
\end{multline}

Because of Eqs. (\ref{eq:oint}), (\ref{eq:F3}), (\ref{eq:F4}) and
(\ref{eq:F3F3F3F4}), it finally demonstrates the orthogonality of
the far-field pattern functions

\begin{multline}
\oiint\oint\mathbf{K}_{smn}(\theta,\phi).\mathbf{K}_{s'm'n'}^{*}(\theta,\phi)d\Omega=\\
4\pi\delta_{ss'}\delta_{mm'}\delta_{nn'}.\label{eq:F4-1-1}
\end{multline}

\subsection{Antenna contribution to the RC $Q$-factor in the lossless case}

\label{subsec:Lossless}

We consider here the case of a lossless system, i.e., $R_{\Omega}=0$
which leads to a radiation efficiency $e_{\mathrm{r}}=1$. Due to
energy conservation, it is possible to write $\sum_{j=0}^{J}\left|S_{ij}\right|^{2}=1\,\forall i$.
By pulling out the term corresponding to $j=0$, it comes $\sum_{j=1}^{J}\left|S_{ij}\right|^{2}=1-\left|S_{i0}\right|^{2}\,\forall i$.
Then, according to the definition of the structual contribution to
the RC $Q$-factor in (\ref{eq:Qstr}), one can show that

\begin{equation}
\frac{Q_{0}}{Q_{\mathrm{s,lossless}}}=\sum_{i=1}^{J}\left|S_{i0}\right|^{2}=e_{\mathrm{r}}=1.
\end{equation}

By taking benefit of the unitary of the $S$-matrix for a lossless
system, the interference mode can be expressed as the following

\begin{multline}
2\Re\left\{ \Gamma{}_{\mathrm{L}}\sum_{i=1}^{J}\left[S_{i0}\sum_{j=1}^{J}\left(S_{0j}S_{ij}^{*}\right)\right]\right\} =\\
2\Re\left\{ \Gamma{}_{\mathrm{L}}\sum_{i=1}^{J}S_{i0}\left[\delta_{\mathrm{0i}}-S_{00}S_{i0}^{*}\right]\right\} \label{eq:F4-1}
\end{multline}

Since $S_{00}=0$, we obtain

\begin{eqnarray}
2\Re\left\{ \Gamma{}_{\mathrm{L}}\sum_{i=1}^{J}\left[S_{i0}\sum_{j=1}^{J}\left(S_{0j}S_{ij}^{*}\right)\right]\right\}  & = & 0.
\end{eqnarray}

Therefore, it is proved here that the model for the antenna contribution
to the RC $Q$-factor in the lossless case is equivalent to the two
models from \cite{Hill1998} and \cite{Cozza2018} in the theoretical
case of $e_{\mathrm{r}}=1$.

\bibliographystyle{IEEEtran}
\addcontentsline{toc}{section}{\refname}\bibliography{MyBib}

\end{document}